\documentclass[aps,prb,twocolumn,floats]{revtex4}
\usepackage{graphics}
\usepackage{amssymb}
\usepackage{amsmath}
\begin{document}

\title{Disorder-induced density of states on the surface of a spherical topological insulator}
\author{Adam C. Durst}
\affiliation{Department of Physics and Astronomy, Hofstra University, Hempstead, NY 11549-1510}
\date{March 24, 2016}

\begin{abstract}
We consider a topological insulator (TI) of spherical geometry and numerically investigate the influence of disorder on the density of surface states.  The energy spectrum of the spherical TI surface is discrete, for a sphere of finite radius, and can be truncated by imposing a high-energy cutoff at the scale of the bulk band gap.  To this clean system we add a surface disorder potential of the most general hermitian form, $V = V^0(\theta,\phi) \openone + {\bf V}(\theta,\phi) \cdot {\bf \sigma}$, where $V^0$ describes the spin-independent part of the disorder and the three components of ${\bf V}$ describe the spin-dependent part.  We expand these four disorder functions in spherical harmonics and draw the expansion coefficients randomly from a four-dimensional, zero-mean gaussian distribution.  Different strengths and classes of disorder are realized by specifying the $4 \times 4$ covariance matrix.  For each instantiation of the disorder, we solve for the energy spectrum via exact diagonalization.  Then we compute the disorder-averaged density of states, $\rho(E)$, by averaging over 200,000 different instantiations.  Disorder broadens the Landau-level delta functions of the clean density of states into peaks that decay and merge together.  If the spin-dependent term is dominant, these peaks split due to the breaking of the degeneracy between time-reversed partner states.  Increasing disorder strength pushes states closer and closer to zero energy (the Dirac point), resulting in a low-energy density of states that becomes nonzero for sufficient disorder, typically approaching an energy-independent saturation value, for most classes of disorder.  But for purely spin-dependent disorder with ${\bf V}$ either entirely out-of-surface or entirely in-surface, we identify intriguing disorder-induced features in the vicinity of the Dirac point.  In the out-of-surface case, a new peak emerges at zero energy.  In the in-surface case, we see a symmetry-protected zero at zero energy, with $\rho(E)$ increasing linearly toward nonzero-energy peaks.  These striking features are explained in terms of the breaking (or not) of two chiral symmetries of the clean Hamiltonian.
\end{abstract}

\maketitle

\section{Introduction}
\label{sec:intro}
Over the past decade, the idea of a topological insulator \cite{has10,qi11,moo10,kan05a,kan05b,fu07a,moo07,roy09,ber06,kon07,fu07b,hsi08,zha09,liu10,xia09} has grown from a theoretical notion to an experimental reality, and has led to the exploration of a wide range of topological quantum phenomena.  Although insulating in the bulk, topological insulators (TIs) are adiabatically distinct from ordinary insulators and support protected gapless surface states, a consequence of the combined effects of spin-orbit interaction and time-reversal symmetry.  In three-dimensional TIs, like Bi$_2$Se$_3$, these surface states exhibit a two-dimensional, spin-momentum-locked, massless Dirac energy spectrum.

Recent work by Neupert {\it et al.\/} \cite{neu15} highlighted the utility of adopting a spherical geometry for numerical studies of the TI surface.  As shown by Imura {\it et al.\/} \cite{imu12}, the massless Dirac Hamiltonian of the flat TI surface can be mapped to a spherical TI surface with the introduction of a fictitious magnetic monopole at the center of the sphere that has opposite sign for electrons of opposite spin.  The finite size of the spherical surface yields a problem with a discrete energy spectrum and well-defined eigenstates that is well-suited to numerical analysis, whether the objective be to study real systems with spherical geometry (i.e.\ TI nanoparticles \cite{imu12}) or to gain insight regarding the flat TI surface that is recovered in the large radius limit.  Neupert {\it et al.\/} \cite{neu15} adopt the latter point of view, making use of spherical geometry to study the effect of electron-electron interactions on TI surface states.

In the present work, we make use of the same spherical geometry \cite{neu15,imu12,gre11,hal83} to perform a numerical disorder analysis, investigating the effects of disorder on the density of states of the TI surface.  Of course, the study of disorder effects in two-dimensional massless Dirac systems has a long history that predates the discovery of topological insulators and was primarily focused on the quasiparticle density of states of so-called dirty $d$-wave superconductors \cite{gor85,lee93,ner95,sen99,dur00,alt02}.  In that system, though the clean superconductor inherits a linear-in-energy quasiparticle density of states from the massless Dirac form of the excitation spectrum, the presence of disorder has the effect of redistributing states from high energy to low.

Here we consider, directly, the case of a disordered TI surface, which differs from the dirty $d$-wave case in that we have a single isotropic Dirac point rather than four anisotropic ones, and most significantly, in that the 2D massless Dirac Hamiltonian acts upon spin-space rather than particle-hole space.  Rather than performing a disorder-averaged perturbative analysis, we make use of the spherical geometry to solve a discrete and truncated problem, via exact diagonalization, for the energy spectrum and density of states for each particular random instantiation of disorder.  We employ a disorder potential of the most general possible form and draw parameters randomly from a four-dimensional, zero-mean gaussian distribution of known covariance matrix.  Results are averaged over 200,000 disorder instantiations to obtain the disorder-averaged density of states as a function of energy.

We begin in Sec.~\ref{ssec:sphere} by reviewing the solution of the clean, single-particle, spherical TI problem, as solved via Refs.~\onlinecite{neu15,imu12,gre11,hal83}, and noting the eigenvalues and eigenstates of the clean Hamiltonian.  In Sec.~\ref{ssec:disorder}, we introduce the disorder potential, parameterize it via a four-component vector-function over the spherical surface, and expand in spherical harmonics.  Then in Sec.~\ref{ssec:random}, we discuss the manner in which those expansion coefficients are randomly selected and the nature of the distribution from which they are drawn.  In Sec.~\ref{ssec:matrixelements}, we make use of the simple form of the eigenstates of the clean system to efficiently compute matrix elements of the disorder potential.  In Sec.~\ref{ssec:diagonalization}, we compute the energy spectrum for each disorder instantiation via exact diagonalization, and in Sec.~\ref{ssec:DOScalc}, we compute the disorder-averaged density of states.  Results for different disorder strengths and different classes of disorder are presented in Sec.~\ref{sec:results} and explained in Sec.~\ref{sec:explanation} in terms of the breaking (or not) of the symmetries of the clean Hamiltonian.  Conclusions are discussed in Sec.~\ref{sec:conclusions}.

\section{Formulation}
\label{sec:formulation}

\subsection{TI surface states on a sphere}
\label{ssec:sphere}
For energies close to the Dirac point, the surface states of a strong three-dimensional topological insulator (TI) are described by the two-dimensional massless Dirac Hamiltonian
\begin{equation}
H_{\rm surf} = v \hat{\bf n} \cdot \left( -i{\bf \nabla} \times {\bf \sigma} \right)
\label{eq:flatDirac}
\end{equation}
where $\hat{\bf n}$ is the surface normal, $v$ is the slope of the Dirac cone, ${\sigma}=(\sigma_1, \sigma_2, \sigma_3)$ is a vector of spin-space Pauli matrices, and energy is measured with respect to the Dirac point.  As shown by Imura {\it et al.\/} \cite{imu12}, mapping this Hamiltonian to a spherical geometry yields
\begin{equation}
H_0 = \frac{v}{R} \left( \sigma_1 \Lambda_\theta + \sigma_2 \Lambda_\phi \right)
\label{eq:sphericalDirac}
\end{equation}
where $R$ is the radius of the sphere, $(r,\theta,\phi)$ are spherical coordinates, and
\begin{equation}
{\bf \Lambda} = -i \left[ \hat{\bf \phi} \frac{\partial}{\partial\theta}
- \hat{\bf \theta} \frac{1}{\sin\theta} \left( \frac{\partial}{\partial\phi} - \frac{i}{2} \cos\theta \sigma_3 \right) \right]
\label{eq:monopole}
\end{equation}
is the dynamical angular momentum of an electron in the presence of a fictitious magnetic monopole of strength $2\pi\sigma_3$ (opposite sign for spin-up versus spin-down) at the center of the sphere.

An elegant solution to this problem, understood in terms of Landau levels on the sphere spanned by two mutually commuting SU(2) algebras (for the cyclotron momentum ${\bf S}$ and the guiding center momentum ${\bf L}$) was obtained by Neupert {\it et al.\/} \cite{neu15} via the formalism developed in Refs.~\onlinecite{gre11,hal83}.  They found exact eigenvalues and eigenvectors of $H_0$, indexed by quantum numbers $n$, $m$, and $\lambda$, where $n=0,1,2,...$ is the Landau level index, $m=-s,-s+1,...,s$ (where $s=n+\frac{1}{2}$) is the eigenvalue of $L_z$, and $\lambda=\pm 1$ for positive/negative energy solutions.  The eigenvalues
\begin{equation}
\epsilon_{nm}^\lambda = \lambda \left( n + 1 \right) \frac{v}{R}
\label{eq:H0eigenvalue}
\end{equation}
are independent of $m$ and therefore $2(n+1)$-fold degenerate.  The spinor eigenstates take the form
\begin{equation}
\psi_{nm}^\lambda (\theta,\phi)
= \left( \begin{array}{c}
\phi_{nm}^\uparrow \\
\lambda \phi_{nm}^\downarrow \end{array} \right)
\label{eq:Psinml}
\end{equation}
with
\begin{equation}
\phi_{nm}^\uparrow = (L^-)^{s-m} \bar{v}^n u^{n+1}
\;\;\;\;\;\;\;\;
\phi_{nm}^\downarrow = -\frac{S^-}{n+1} \phi_{nm}^\uparrow
\label{eq:phiupdown}
\end{equation}
where the $L^-$ and $S^-$ operators
\begin{equation}
L^- \equiv v \partial_u - \bar{u} \partial_{\bar{v}}
\;\;\;\;\;\;\;\;
S^- \equiv \bar{v} \partial_u - \bar{u} \partial_v
\label{eq:LminusSminus}
\end{equation}
are defined in terms of the spinor coordinates
\begin{equation}
u \equiv \cos(\theta/2) e^{i\phi/2}
\;\;\;\;\;\;\;\;
v \equiv \sin(\theta/2) e^{-i\phi/2}
\label{eq:uvdef}
\end{equation}
introduced by Haldane \cite{hal83}.  The eigenstates are orthogonal and we normalize over the unit sphere, setting
\begin{equation}
\left\langle \psi_{nm}^\lambda | \psi_{nm}^\lambda \right\rangle
= \int_0^{2\pi} \!\!\!\!\!\!d\phi \int_0^\pi \!\!\!\!d\theta \sin\theta
\left[ \left| \phi_{nm}^\uparrow \right|^2 + \left| \phi_{nm}^\downarrow \right|^2 \right] = 1 .
\label{eq:normalization}
\end{equation}

As discussed by Neupert {\it et al.\/} \cite{neu15}, the weight of these surface states, valid in the vicinity of the Dirac point, diminishes as they merge with the bulk conduction and valence bands for positive and negative energies on the scale of the bulk energy gap.  Noting that even strong electron-electron interactions would therefore only induce small matrix elements between surface and bulk states, they neglected such matrix elements and confined their analysis to surface states within $\Lambda$ of the Dirac point, where $\Lambda$ is a cutoff energy on the scale of the bulk energy gap.  We take the same approach here, neglecting disorder matrix elements between surface and bulk states and restricting our Hilbert space to surface states with $|\epsilon_{nm}^\lambda| < \Lambda$.

Since the energy spacing between Landau levels is $v/R$, this restriction limits the energy spectrum to a finite number of states, $N$, with that number controlled by the radius of the sphere.  The form and degeneracy of the energy spectrum, Eq.~(\ref{eq:H0eigenvalue}), reveal that
\begin{equation}
N = 2 \left( n_{\rm max} + 1 \right) \left( n_{\rm max} + 2 \right)
\label{eq:numberofstates}
\end{equation}
where
\begin{equation}
n_{\rm max} = \left\lfloor \frac{R\Lambda}{v} - 1 \right\rfloor
\label{eq:maxLandaulevel}
\end{equation}
is the maximum Landau level index (the brackets denote the floor or greatest integer function).  With these $N$ surface states of the clean system in hand, our task will be to define the disorder potential, calculate the $N \times N$ matrix elements of the full Hamiltonian (including disorder), and solve for energy spectra via exact diagonalization.

\subsection{Disorder potential}
\label{ssec:disorder}
The presence of disorder will mix the eigenstates of the clean TI surface.  We consider a disorder potential operator of the most general possible form, a $2 \times 2$ (in spin-space) hermitian matrix-function defined over the spherical surface.  Such a potential can be written as
\begin{equation}
V(\theta,\phi) = \sum_{\alpha=0}^3 V^\alpha(\theta,\phi) \,\sigma_\alpha = V^0 \openone + {\bf V} \cdot {\bf \sigma}
\label{eq:Vdef}
\end{equation}
where $\sigma_1$, $\sigma_2$, and $\sigma_3$ are spin-space Pauli matrices, $\sigma_0 \equiv \openone$ is the identity matrix, and the $V^\alpha(\theta,\phi)$ are four real functions of angular variables $\theta$ and $\phi$.  The first equality in Eq.~(\ref{eq:Vdef}) suggests we interpret the $V^\alpha$ as components of a four-vector in $\mathbb{R}^4$, while the second equality further identifies the last three components as the three-vector ${\bf V}$.  Note that the $V^0\openone$ term commutes with the time-reversal operator, $T \equiv -i\sigma_2K$ (where K denotes complex conjugation), while the ${\bf V} \cdot {\bf \sigma}$ term anticommutes with $T$.  Thus, $V^0$ characterizes the time-reversal-invariant, spin-independent part of the disorder (i.e.\ due to nonmagnetic impurities) while ${\bf V}$ characterizes the time-reversal-breaking, spin-dependent part (i.e.\ due to magnetic impurities).  In what follows, it will prove important to make note of the orientation of ${\bf V}$ and distinguish between the in-surface part (perpendicular to the local surface normal) and the out-of-surface part (parallel to the local surface normal) as they have remarkably different effects on the low-energy density of states.

We expand each of these four functions in spherical harmonics
\begin{equation}
V^\alpha(\theta,\phi) = \sum_{\ell=0}^\infty \sum_{m=-\ell}^\ell V^\alpha_{\ell m} Y_\ell^m(\theta,\phi) ,
\label{eq:Ylmexpansion}
\end{equation}
restricting the expansion coefficients such that the sums are real.  Noting that $Y_\ell^{m *} = (-1)^m Y_\ell^{-m}$, this restriction is imposed by requiring that
\begin{equation}
V_{\ell,-m}^\alpha = (-1)^m V_{\ell m}^{\alpha *}
\label{eq:coefcondition}
\end{equation}
which, for $m=0$, requires that the $V^\alpha_{\ell 0}$ be real.  It is these coefficients that will be randomly selected to determine an instantiation of the disorder potential.

As noted in Ref.~\onlinecite{neu15}, the high-energy cutoff at $\Lambda$ means that it is not possible, within our restricted Hilbert space, to construct orbitals that are localized in position space on length scales smaller than $2\pi v/\Lambda$.  Thus, short-wavelength disorder on length scales shorter than this is simply averaged over, so we need only include in our expansion spherical harmonics of angular momentum up to $R\Lambda/v$.  Hence, in our numerics, we shall truncate the sum over $\ell$ in Eq.~(\ref{eq:Ylmexpansion}) at $\ell_{\rm max} \equiv \lfloor R\Lambda/v \rfloor = n_{\rm max} + 1$.

\section{Numerical Methods}
\label{sec:numericalmethods}

\subsection{Random draw of disorder potential coefficients}
\label{ssec:random}
A particular instantiation of disorder is simulated by randomly selecting values for each of the $V^\alpha_{\ell m}$ disorder coefficients defined in Sec.~\ref{ssec:disorder}.  In light of the restrictions imposed by Eq.~(\ref{eq:coefcondition}), we require one real random four-vector $V$ per $\ell$-$m$ pair.  Each is drawn from a four-dimensional gaussian distribution
\begin{equation}
p(V;\Sigma) = \frac{1}{(2\pi)^2\sqrt{\det \Sigma}} \exp \left( -\frac{1}{2} V^T \Sigma^{-1} V \right)
\label{eq:gaussian}
\end{equation}
of zero mean and $4 \times 4$ covariance matrix $\Sigma$.  The covariance matrix is the means by which we control the strength and type of disorder simulated.  The shape of the one-sigma hyper-ellipsoid determines the relative magnitude of the terms in Eq.~(\ref{eq:Vdef}).  For our current purposes, it has been sufficient to consider a diagonal covariance matrix
\begin{equation}
\Sigma = \left[ \begin{array}{cccc}
s_0^2 & 0 & 0 & 0 \\
0 & s_1^2 & 0 & 0 \\
0 & 0 & s_2^2 & 0 \\
0 & 0 & 0 & s_3^2 \end{array} \right]
\label{eq:diagonalSigma}
\end{equation}
such that the probability distribution reduces to the product of four independent one-dimensional gaussians
\begin{equation}
p(V;\Sigma) = \prod_{\alpha=0}^3 \frac{1}{\sqrt{2\pi} s_\alpha} \exp \left( -\frac{V_\alpha^2}{2 s_\alpha^2} \right)
\label{eq:diagonalgaussian}
\end{equation}
where we have used $s_\alpha$ (rather than $\sigma_\alpha$) to denote the standard deviation along each principal axis (to avoid notational confusion with the Pauli matrices).

In what follows, we shall use the root-mean-square (rms) value of the magnitude of $V$ as our measure of the strength of the disorder.  To separate strength from type of disorder, we define a disorder strength parameter, $V_{\rm scale}$, and set it equal to $\sqrt{\langle V^2 \rangle}$, which necessarily constrains the $s_\alpha$ since
\begin{eqnarray}
V_{\rm scale}^2 &\equiv& \langle V^2 \rangle = \int d^4V \left( V_0^2 + V_1^2 + V_2^2 + V_3^2 \right) p(V;\Sigma) \nonumber \\
&=& s_0^2 + s_1^2 + s_1^2 + s_3^2 .
\label{eq:sigmaconstraint}
\end{eqnarray}
To enforce this constraint, we set
\begin{equation}
s_\alpha = V_{\rm scale} \frac{\beta_\alpha}{\sqrt{\beta_0^2 + \beta_1^2 + \beta_2^2 + \beta_3^2}}
\label{eq:constrainedsigmas}
\end{equation}
where $\beta$ is a four-vector that specifies the type of disorder, independent of the disorder strength.  For example, $\beta = [1\ 0\ 0\ 0]$ specifies time-reversal-invariant disorder, while $\beta = [0\ 1\ 1\ 1]$ specifies time-reversal-breaking disorder drawn from a uniform mix of ${\bf V}$ orientations, and $\beta = [0\ 1\ 1\ 0]$ specifies in-surface time-reversal-breaking disorder.  For any choice of $\beta$-vector, we vary the disorder strength by changing $V_{\rm scale}$.

\begin{figure}
\centerline{\resizebox{3.25in}{!}{\includegraphics{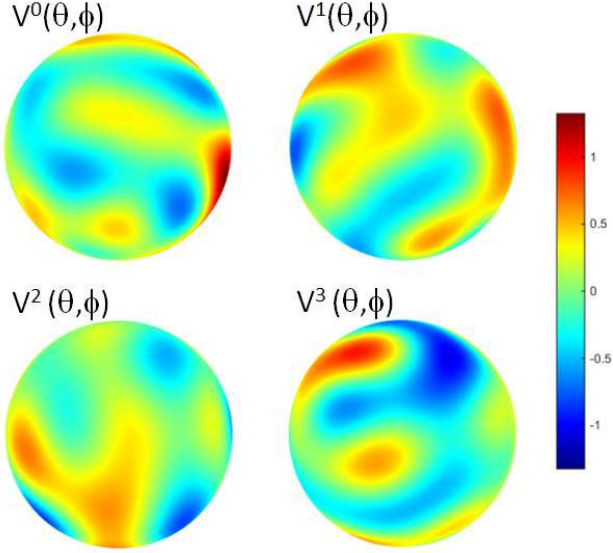}}}
\caption{(Color online) Sample disorder instantiation.  Displayed are the components, $V^\alpha(\theta,\phi)$, of the four-vector function parameterizing the disorder potential operator, plotted over the spherical surface.  These patterns depict one particular instantiation of the disorder potential, randomly drawn from a gaussian distribution with a covariance matrix characterized by $\beta = [1\ 1\ 1\ 1]$ and $V_{\rm scale}=0.4 v/R$.  The color bar indicates disorder intensity in units of $v/R$.}
\label{fig:disorderinstantiation}
\end{figure}

In general, one could specify a different covariance matrix ($\beta$-vector and $V_{\rm scale}$) for every $\ell$ and $m$, to tailor the length-scale-mix of the simulated disorder.  For simplicity, in this analysis, we have taken the covariance matrix to be uniform across all the $\ell$ and $m$ for which $\ell \leq \ell_{\rm max}$. Displayed in Fig.~\ref{fig:disorderinstantiation} are the four components, $V^\alpha(\theta,\phi)$, of a sample disorder instantiation drawn from a distribution with $\beta = [1\ 1\ 1\ 1]$ and $V_{\rm scale}=0.4 v/R$ and plotted over the spherical surface.

\subsection{Matrix element computation}
\label{ssec:matrixelements}
With the $2 \times 2$ disorder potential operator, $V(\theta,\phi)$, in hand, our task is to compute the $N \times N$ matrix elements of this operator between the $N$ eigenstates of the clean Hamiltonian.  Each of these matrix elements takes the form
\begin{equation}
V_{ij} \equiv V(n_i,m_i,\lambda_i;n_j,m_j,\lambda_j) = \langle \psi_{n_i m_i}^{\lambda_i} | V(\theta,\phi) | \psi_{n_j m_j}^{\lambda_j} \rangle
\label{eq:Vijdef}
\end{equation}
indexed by the quantum numbers of the bra eigenstate, $(n_i,m_i,\lambda_i)$ and those of the ket eigenstate $(n_j,m_j,\lambda_j)$.  Combining Eqs.~(\ref{eq:Vdef}) and (\ref{eq:Ylmexpansion}) and exchanging the order of the summations, we can write the disorder potential operator as
\begin{equation}
V(\theta,\phi) = \sum_{\ell=0}^{\ell_{\rm max}} \sum_{m=-\ell}^\ell V_{\ell m} Y_\ell^m(\theta,\phi)
\label{eq:Vexpansion}
\end{equation}
where $V_{\ell m}$ is the $2 \times 2$ expansion coefficient
\begin{equation}
V_{\ell m} \equiv \sum_{\alpha=0}^3 V_{\ell m}^\alpha \sigma_\alpha
= \left[ \begin{array}{cc}
V_{\ell m}^{\uparrow\uparrow} & V_{\ell m}^{\uparrow\downarrow} \\
V_{\ell m}^{\downarrow\uparrow} & V_{\ell m}^{\downarrow\downarrow} \end{array} \right] .
\label{eq:Vlm}
\end{equation}
Plugging Eqs.~(\ref{eq:Psinml}), (\ref{eq:Vexpansion}), and (\ref{eq:Vlm}) into Eq.~(\ref{eq:Vijdef}), we find that the matrix element takes the form
\begin{eqnarray}
V_{ij} = \sum_{\ell=0}^{\ell_{\rm max}} \sum_{m=-\ell}^\ell &\Big[&
V_{\ell m}^{\uparrow\uparrow} M_{ij;\ell m}^{\uparrow\uparrow}
+ V_{\ell m}^{\uparrow\downarrow} M_{ij;\ell m}^{\uparrow\downarrow} \nonumber \\
&+& V_{\ell m}^{\downarrow\uparrow} M_{ij;\ell m}^{\downarrow\uparrow}
+ V_{\ell m}^{\downarrow\downarrow} M_{ij;\ell m}^{\downarrow\downarrow} \Big]
\label{eq:VijFromMij}
\end{eqnarray}
where
\begin{eqnarray}
M_{ij;\ell m}^{\uparrow\uparrow} &\equiv& \langle \phi_{n_i m_i}^\uparrow | Y_\ell^m | \phi_{n_j m_j}^\uparrow \rangle \nonumber \\
M_{ij;\ell m}^{\uparrow\downarrow} &\equiv& \lambda_j \langle \phi_{n_i m_i}^\uparrow | Y_\ell^m | \phi_{n_j m_j}^\downarrow \rangle \nonumber \\
M_{ij;\ell m}^{\downarrow\uparrow} &\equiv& \lambda_i \langle \phi_{n_i m_i}^\downarrow | Y_\ell^m | \phi_{n_j m_j}^\uparrow \rangle \nonumber \\
M_{ij;\ell m}^{\downarrow\downarrow} &\equiv& \lambda_i \lambda_j \langle \phi_{n_i m_i}^\downarrow | Y_\ell^m | \phi_{n_j m_j}^\downarrow \rangle
\label{eq:Mijdef}
\end{eqnarray}
and the inner product is defined over the unit sphere
\begin{equation}
\langle f | g \rangle \equiv \int_0^{2\pi} \!\!\!\!\!\!d\phi \int_0^\pi \!\!\!\!d\theta \sin\theta f(\theta,\phi)^* g(\theta,\phi) .
\label{eq:innerproduct}
\end{equation}

Note that it is critical that we have separated the $M_{ij;\ell m}$ factors, which are computationally expensive to calculate but are the same for every disorder instantiation, from the $V_{\ell m}$ factors, which can be obtained quickly but must be drawn anew for every disorder instantiation.  We computed and stored the $M_{ij;\ell m}$ factors only once, reusing them for each of the millions of disorder instantiations that were run.

To calculate the inner products in the $M_{ij;\ell m}$ factors, we take advantage of the fact that the eigenstates of the clean Hamiltonian as well as the spherical harmonic functions can all be expressed as finite polynomials (with complex $A_i$ and integer $p_i$, $q_i$, and $r_i$) of the form
\begin{equation}
\sum_i A_i \left( \sin \frac{\theta}{2} \right)^{p_i} \left( \cos \frac{\theta}{2} \right)^{q_i} \left( e^{i\phi/2} \right)^{r_i}
\label{eq:SCEpolydef}
\end{equation}
which we refer to as SCE (sin-cos-exp) polynominals.  For the eigenstates, this follows directly from the form of Eqs.~(\ref{eq:phiupdown}) through (\ref{eq:uvdef}).  For the spherical harmonics, it follows from expanding the associated Legendre functions \cite{gra94} and writing the sines and cosines in terms of half-angle sines and cosines.  Doing so, we find that
\begin{equation}
Y_\ell^m = \sum_{k=0}^{\left\lfloor \frac{\ell-|m|}{2} \right\rfloor} \sum_{j=0}^{\ell-2k-|m|} C_{\ell mkj}
\left( \sin \frac{\theta}{2} \right)^{p} \left( \cos \frac{\theta}{2} \right)^{q} \left( e^{i\phi/2} \right)^{r}
\label{eq:YlmSCE}
\end{equation}
where $p=2j+|m|$, $q=2(\ell-2k-|m|-j)+|m|$, $r=2m$,
\begin{eqnarray}
C_{\ell mkj} &\equiv& A_\ell^m 2^{|m|-\ell} (-1)^{k+j} |m|!
\left( \begin{array}{c} \ell \\ k \end{array} \right)
\left( \begin{array}{c} 2\ell-2k \\ \ell \end{array} \right) \nonumber \\
&\times& \left( \begin{array}{c} \ell-2k \\ |m| \end{array} \right)
\left( \begin{array}{c} \ell-2k-|m| \\ j \end{array} \right)
\label{eq:Clmkj}
\end{eqnarray}
\begin{equation}
A_\ell^m \equiv \sqrt{\frac{2\ell+1}{4\pi} \frac{(\ell-|m|)!}{(\ell+|m|)!}} \times
\left\{ \begin{array}{r} (-1)^m \;\;\mbox{for}\;\; m \geq 0 \\ 1 \;\;\;\;\;\;\mbox{for}\;\; m < 0 \end{array} \right.
\label{eq:Alm}
\end{equation}
and the parentheses denote binomial coefficients.  Note that the set of SCE polynomials is closed under addition, multiplication, and complex conjugation.  Thus, since the measure in our inner product, $\sin\theta = 2\sin\frac{\theta}{2}\cos\frac{\theta}{2}$, is also an SCE polynomial, evaluating the inner products in the $M_{ij;\ell m}$ factors amounts to integrating the terms of an SCE polynomial over $\theta$ and $\phi$.  This can always be done analytically since
\begin{eqnarray}
\lefteqn{\int_0^{2\pi} \!\!\!\!\!\!d\phi \int_0^\pi \!\!\!\!d\theta
\left( \sin \frac{\theta}{2} \right)^{p} \left( \cos \frac{\theta}{2} \right)^{q} \left( e^{i\phi/2} \right)^{r}} \\
&& = B \left( \frac{p+1}{2},\frac{q+1}{2} \right) \times
\left\{ \begin{array}{l} 2\pi \;\;\mbox{for}\;\; r = 0 \\
4i/r \;\;\mbox{for}\;\; r \neq 0, \;\mbox{odd} \\
0  \;\;\mbox{for}\;\; r \neq 0, \;\mbox{even}\end{array} \right. \nonumber
\label{eq:SCEintegral}
\end{eqnarray}
for $p,q>-1$ (a condition that is always met for us), where $B(x,y) \equiv \Gamma(x)\Gamma(y)/\Gamma(x+y)$ is a Beta function \cite{gra94}.  Thus, our numerics need only keep track of the various terms and add them in order to compute the $M_{ij;\ell m}$ factors, and this need only be done once.  Then, for each instantiation of the disorder, a new set of $V_{\ell m}$ coefficients are drawn at random from the desired distribution (as per Sec.~\ref{ssec:random}) and plugged into Eq.~(\ref{eq:VijFromMij}) to obtain the matrix elements of the disorder potential operator.

\subsection{Exact diagonalization}
\label{ssec:diagonalization}
\begin{figure}
\centerline{\resizebox{3in}{!}{\includegraphics{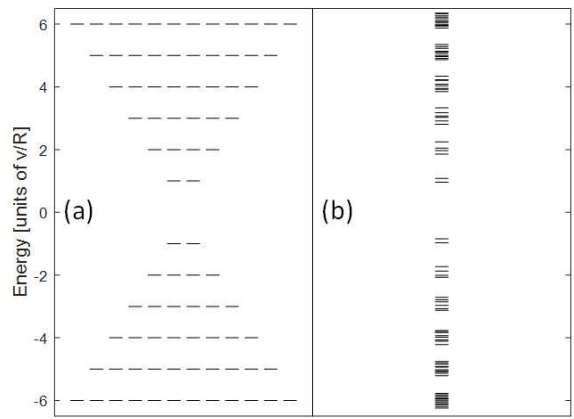}}}
\caption{Energy spectra of the spherical TI surface calculated via exact diagonalization up to $n_{\rm max}=5$ ($\Lambda=6.5v/R$, $N=84$ states).  Plotted in (a) are the 84 energy levels of the clean system, with degenerate levels shown side by side.  Plotted in (b) is the corresponding energy spectrum in the presence of the disorder instantiation depicted in Fig.~\ref{fig:disorderinstantiation}, a particular random draw from a distribution of type $\beta = [1\ 1\ 1\ 1]$ and strength $V_{\rm scale}=0.4 v/R$.  For this disorder type and strength, the degeneracy of each Landau level has been broken, with its states spread out in energy but not yet overlapping with those of neighboring Landau levels.}
\label{fig:spectra}
\end{figure}

In the basis of the $N$ eigenstates of the clean Hamiltonian, the $N \times N$ disordered Hamiltonian matrix is given by
\begin{equation}
{\cal H} = {\cal H}_0 + {\cal V}
\label{eq:Hdisordered}
\end{equation}
where ${\cal H}_0$ is a diagonal matrix of the clean eigenvalues, $\epsilon_{nm}^\lambda$, given in Eq.~(\ref{eq:H0eigenvalue}), and ${\cal V}$ is one instantiation of the disorder potential matrix computed as per Secs.~\ref{ssec:random} and \ref{ssec:matrixelements}.  For each instantiation of the disorder, we numerically compute the eigenvalues of ${\cal H}$ to obtain the $N$ energy levels of that particular instantiation of the disordered system.  Plotted in Fig.~\ref{fig:spectra} is the energy spectrum calculated for the disorder instantiation shown in Fig.~\ref{fig:disorderinstantiation}, displayed alongside the energy spectrum of the clean system.  Note how the degeneracy of each Landau level has been broken by the disorder, spreading out the distribution of energy levels.

\subsection{Disorder-averaged density of states}
\label{ssec:DOScalc}
Though each instantiation of the disorder generates a distinct and discrete spectrum of energy levels, the disorder-averaged density of states is a continuous function of energy that reflects the properties of the entire distribution of disorder potentials, rather than any particular instantiation.  We define
\begin{equation}
\rho(E) \equiv \left\langle \frac{\Delta N(E,\Delta E)}{\Delta E} \right\rangle_{\rm disorder}
\label{eq:DADOS}
\end{equation}
where $\Delta N(E,\Delta E)$ is the number of levels within an energy range of width $\Delta E$ about energy $E$, and the angle brackets denote an average over disorder instantiations.  For a particular disorder instantiation, the function $\Delta N(E,\Delta E)$ is easily identified as a histogram of energy levels of bin size $\Delta E$, and is computed by sorting our energy levels into bins.  Thus, our procedure to compute $\rho(E)$ is to calculate the energy spectrum for a given disorder draw, sort into bins, divide by the bin size, and repeat, averaging over many instantiations of disorder.

Ideally, we would average over an infinite number of instantiations and take the limit of the above as $\Delta E \rightarrow 0$.  Limited, however, by the constraints of our computing power, we settle for averaging over as many instantiations as is computationally possible.  In this analysis, we have computed 200,000 instantiations per run.  With this number finite, there are trade-offs to consider in selecting a bin size.  Choose the bin size too large and lose energy resolution in computing $\rho(E)$.  Choose it too small, and there is insufficient data to beat down the noise about the disorder average.  We found $\Delta E = 0.01 v/R$ to be a reasonable compromise and have used it throughout this analysis.

\begin{figure}[!ht]
\centerline{\resizebox{3in}{!}{\includegraphics{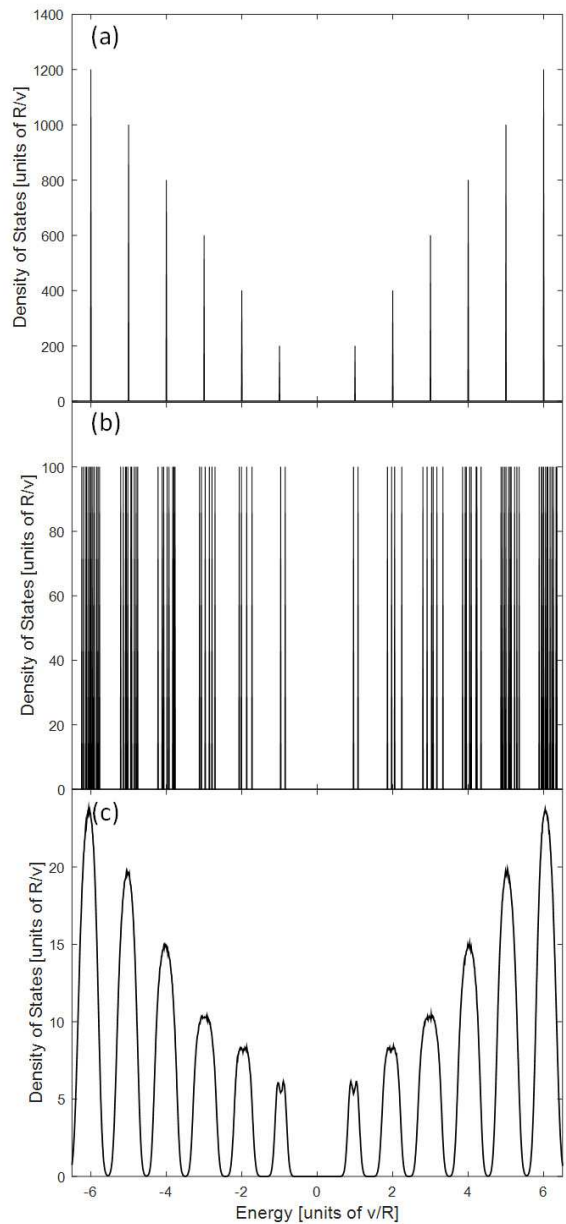}}}
\caption{Density of states as a function of energy for (a) the clean system, (b) a particular instantiation of the disorder, and (c) a disorder average over 200,000 different instantiations.  All instantiations included in this disorder average were drawn from a distribution of type $\beta = [1\ 1\ 1\ 1]$ with strength $V_{\rm scale}=0.4 v/R$.  The particular disorder instantiation used to create (b) is the one plotted on the spherical surface in Fig.~\ref{fig:disorderinstantiation} and for which the energy spectrum in Fig.~\ref{fig:spectra}(b) was calculated.  Results were computed via exact diagonalization up to $n_{\rm max}=5$ ($\Lambda=6.5v/R$, $N=84$ states).}
\label{fig:dosdemo}
\end{figure}

Plotted in Fig.~\ref{fig:dosdemo}(a) is the density of states for the clean system, obtained from the clean energy spectrum in Fig.~\ref{fig:spectra}(a).  As expected, states are confined to the Landau levels, with the Landau level degeneracy growing linearly with $|E|$, as must be the case for a massless Dirac system.  Plotted in Fig.~\ref{fig:dosdemo}(b) is the density of states for a single instantiation of disorder, the one shown in Fig.~\ref{fig:disorderinstantiation}.  The effect of this disorder is to break the degeneracy of the Landau levels, spreading out the states as per the energy spectrum of Fig.~\ref{fig:spectra}(b).  For each randomly drawn disorder instantiation, a slightly different set of levels is obtained.  Averaging over 200,000 instantiations yields the disorder-averaged density of states, as plotted in Fig.~\ref{fig:dosdemo}(c).

\section{Results}
\label{sec:results}
Using the methods described above, we have computed the disorder-averaged density of states, $\rho(E)$, as a function of energy, $E$, and disorder strength, $V_{\rm scale}$, for several important classes of disorder, as specified by the $\beta$-vectors introduced in Sec.~\ref{ssec:random}.  All calculations were performed via exact diagonalization up to $n_{\rm max}=5$ ($\Lambda=6.5v/R$, $N=84$ states) and averaged over 200,000 disorder instantiations per run.  Density of states was computed with an energy bin size of $\Delta E = 0.01 v/R$.

\begin{figure}
\centerline{\resizebox{3in}{!}{\includegraphics{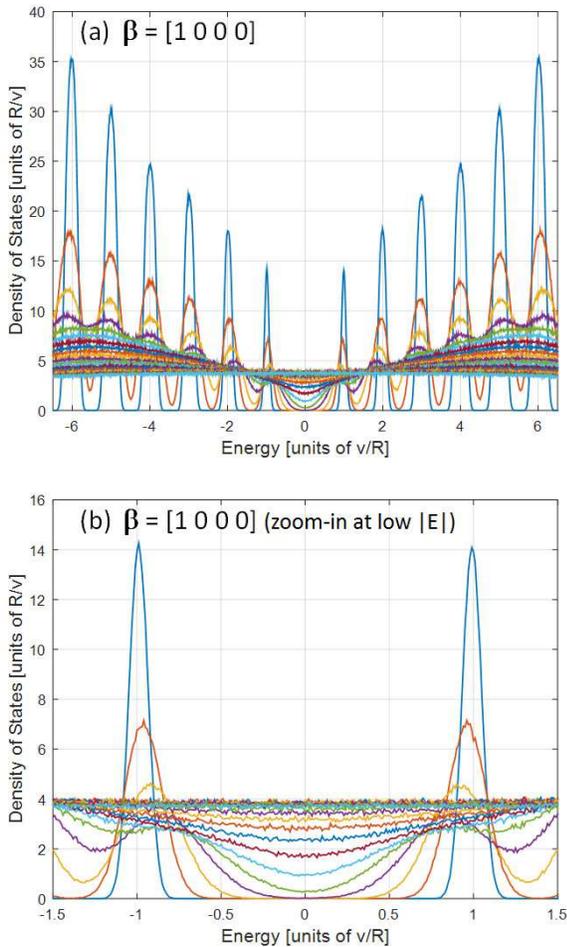}}}
\caption{(Color online) Disorder-averaged density of states for a spherical TI surface in the presence of disorder of type $\beta = [1\ 0\ 0\ 0]$, time-reversal-invariant (spin-independent) disorder, of nonzero variance only in the $V^0$ direction.  Results for twenty values of disorder strength are plotted in (a), from $V_{\rm scale}=0.2 v/R$ through $4 v/R$ in steps of $0.2 v/R$ (bottom to top at $E=0$).  A zoomed-in view of the same data, for $|E| < 1.5 v/R$, is shown in (b).}
\label{fig:dadosS}
\end{figure}

We begin with the case of disorder drawn from a gaussian distribution that is of zero variance in all dimensions but the first, the time-reversal-invariant (spin-independent) case where $V = V^0 \openone$.  In our $\beta$-vector nomenclature, this is $\beta=[1\ 0\ 0\ 0]$.  Results for twenty values of disorder strength, up to $V_{\rm scale}=4v/R$, are plotted in Fig.~\ref{fig:dadosS}.  With increasing disorder strength, the delta functions of the clean density of states become finite peaks, which diminish and broaden until they all overlap.  The regions between delta functions, including the vicinity of zero energy (the Dirac point), thereby fill in with states.  For sufficient disorder, $\rho(0)$ attains a nonzero value, which grows with disorder until saturating.

\begin{figure}
\centerline{\resizebox{3in}{!}{\includegraphics{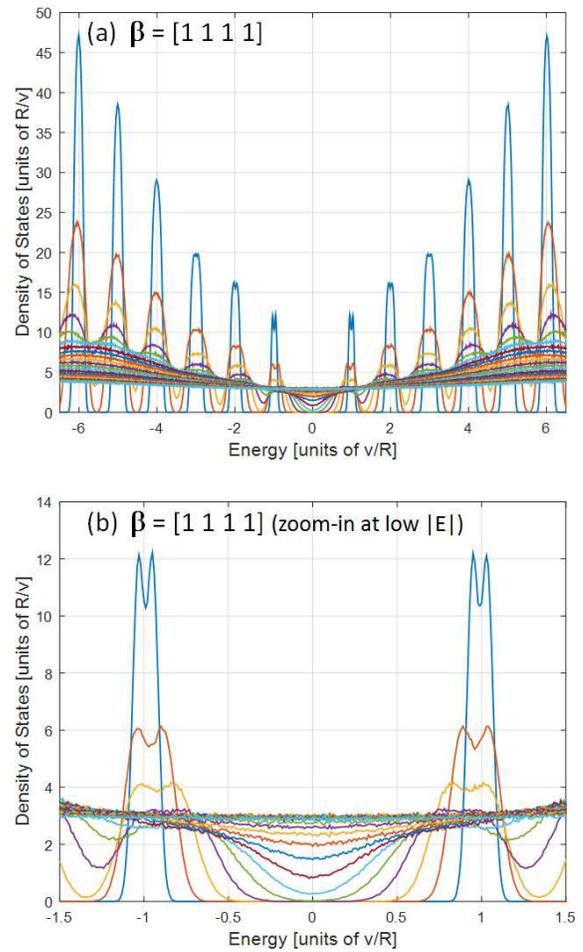}}}
\caption{(Color online) Disorder-averaged density of states for a spherical TI surface in the presence of disorder of type $\beta = [1\ 1\ 1\ 1]$, a mixture of time-reversal-invariant (spin-independent) and time-reversal-breaking (spin-dependent) disorder, of (equal) nonzero variance in the $V^0$, $V^1$, $V^2$, and $V^3$ directions.  Results for twenty values of disorder strength are plotted in (a), from $V_{\rm scale}=0.2 v/R$ through $4 v/R$ in steps of $0.2 v/R$ (bottom to top at $E=0$).  A zoomed-in view of the same data, for $|E| < 1.5 v/R$, is shown in (b).}
\label{fig:dadosSXYZ}
\end{figure}

Now consider the case where all four terms of the disorder potential, $V = V^0 \openone + V^1 \sigma_1 + V^2 \sigma_2 + V^3 \sigma_3$, are treated on equal footing, drawn from a symmetrical four-dimensional gaussian distribution, a mixture of time-reversal-invariant (spin-independent) and time-reversal-breaking (spin-dependent) disorder.  This is the $\beta=[1\ 1\ 1\ 1]$ case, results for which are plotted in Fig.~\ref{fig:dadosSXYZ}.  Once again, with increasing disorder strength, the peaks in $\rho(E)$ at each Landau level diminish, broaden, and eventually overlap.  And once again, $\rho(0)$ attains a nonzero value for sufficient disorder strength.  Notice however, that in this case, there is a small splitting in the tops of the weak-disorder peaks, especially in the lowest energy peaks (those descended from the lowest index Landau levels).

\begin{figure}
\centerline{\resizebox{3in}{!}{\includegraphics{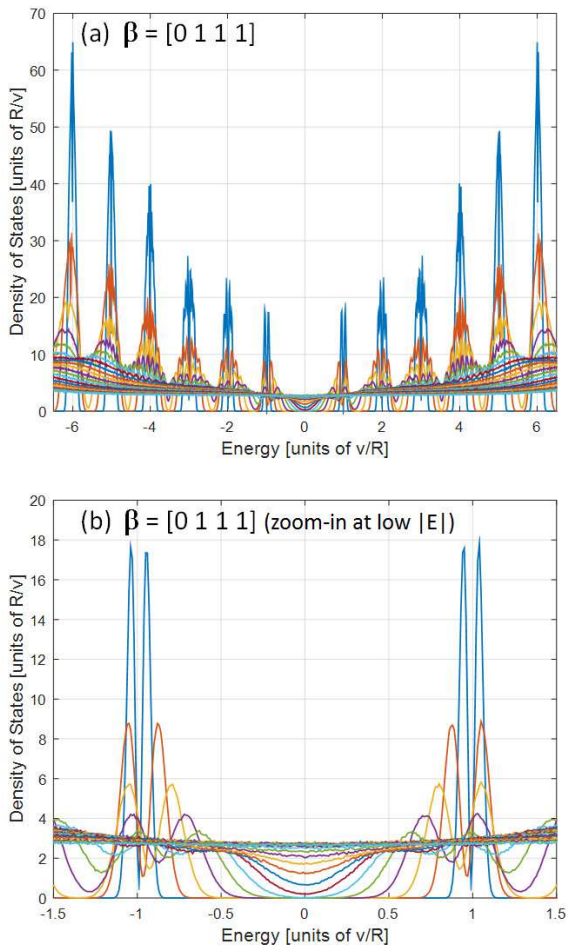}}}
\caption{(Color online) Disorder-averaged density of states for a spherical TI surface in the presence of disorder of type $\beta = [0\ 1\ 1\ 1]$, time-reversal-breaking (spin-dependent) disorder, of (equal) nonzero variance only in the $V^1$, $V^2$, and $V^3$ directions.  Results for twenty values of disorder strength are plotted in (a), from $V_{\rm scale}=0.2 v/R$ through $4 v/R$ in steps of $0.2 v/R$ (bottom to top at $E=0$).  A zoomed-in view of the same data, for $|E| < 1.5 v/R$, is shown in (b).}
\label{fig:dadosXYZ}
\end{figure}

To explore this effect further, let us consider disorder of type $\beta = [0\ 1\ 1\ 1]$, purely time-reversal-breaking (spin-dependent) disorder.  Here the disorder potential has the form $V = V^1 \sigma_1 + V^2 \sigma_2 + V^3 \sigma_3 = {\bf V} \cdot {\bf \sigma}$, with coefficients drawn from a gaussian distribution with equal variance in the three components of ${\bf V}$ and no $V^0$ term at all.  Results are plotted in Fig.~\ref{fig:dadosXYZ}.  With the time-reversal-invariant term missing, the weak-disorder peak splitting is much more pronounced and occurs in every Landau level.  This peak splitting begins upon inclusion of even very weak disorder and persists until the peaks themselves have flattened out.  The source of this effect, and why it does not occur for $\beta = [1\ 0\ 0\ 0]$, is discussed in Sec.~\ref{sec:explanation}.

\begin{figure}
\centerline{\resizebox{3in}{!}{\includegraphics{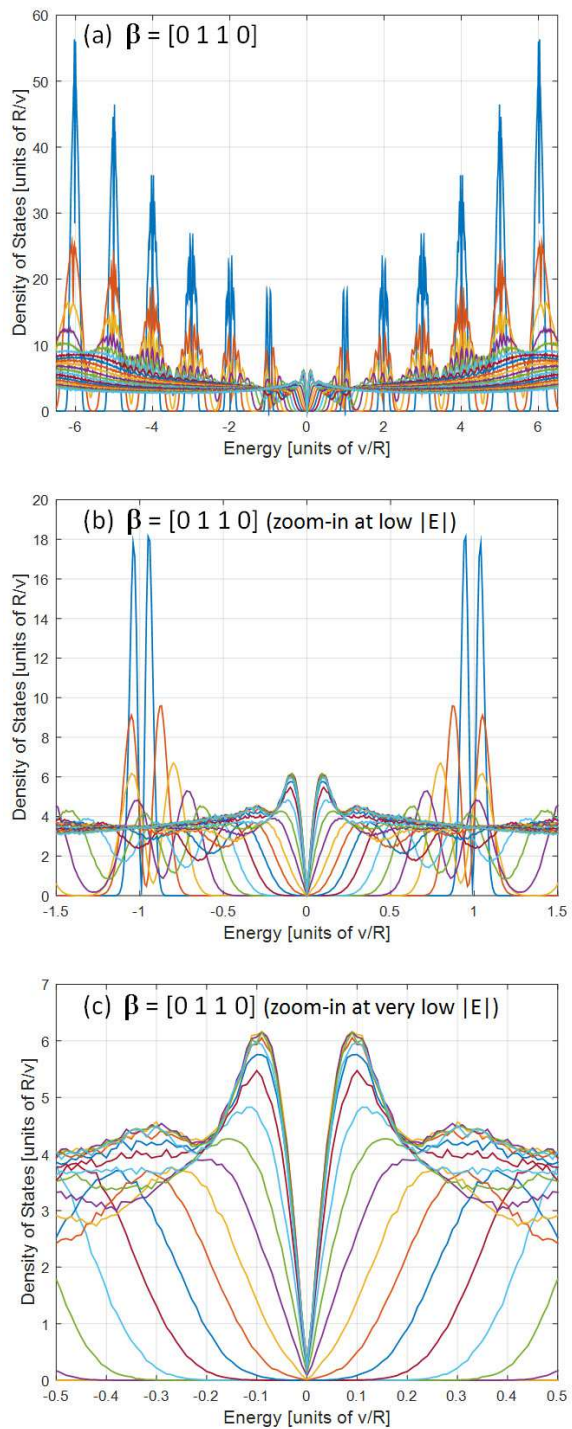}}}
\caption{(Color online) Disorder-averaged density of states for a spherical TI surface in the presence of disorder of type $\beta = [0\ 1\ 1\ 0]$, in-surface time-reversal-breaking (spin-dependent) disorder, of (equal) nonzero variance only in the $V^1$ and $V^2$ directions.  Results for twenty values of disorder strength are plotted in (a), from $V_{\rm scale}=0.2 v/R$ through $4 v/R$ in steps of $0.2 v/R$ (bottom to top at $E=\pm 0.1 v/R$).  A zoomed-in view of the same data, for $|E| < 1.5 v/R$, is shown in (b), and a further zoomed-in view, for $|E| < 0.5 v/R$, is shown in (c).}
\label{fig:dadosXY}
\end{figure}

Now consider what happens when we restrict the time-reversal-breaking disorder vector, ${\bf V}$, to be perpendicular to the surface normal, such that the disorder potential is of the form $V = V^1 \sigma_1 + V^2 \sigma_2$, with these two in-surface components drawn with equal variance and the $V^0$ and $V^3$ terms set to zero.  Results for this $\beta = [0\ 1\ 1\ 0]$ case are plotted in Fig.~\ref{fig:dadosXY}.  As before, peaks in $\rho(E)$ diminish, broaden, and overlap as disorder strength increases.  And as in the prior case, we observe peak splitting for weak disorder.  But here we see something new and quite surprising in the vicinity of the Dirac point.  In all the previous cases, as the density of states filled in between the two lowest $|E|$ peaks, the zero-energy density of states, $\rho(0)$, attained a nonzero value for sufficient disorder strength and continued to grow until saturation at a constant level.  But in the present case, the zero-energy density of states is always zero.  As disorder strength increases and states shift to lower energies, the zero at $E=0$ is preserved, with the density of states building up in two new peaks at $E \approx \pm 0.1 v/R$ and then decreasing linearly to zero as $|E|$ approaches zero ($\rho(E) \sim |E|$ for small $|E|$).  This feature in the low-energy density of states for in-surface, time-reversal-breaking disorder, which we zoom in upon in Fig.~\ref{fig:dadosXY}(c), is a quite striking result.  A discussion of the symmetry that preserves it is presented in Sec.~\ref{sec:explanation}.

\begin{figure}
\centerline{\resizebox{3in}{!}{\includegraphics{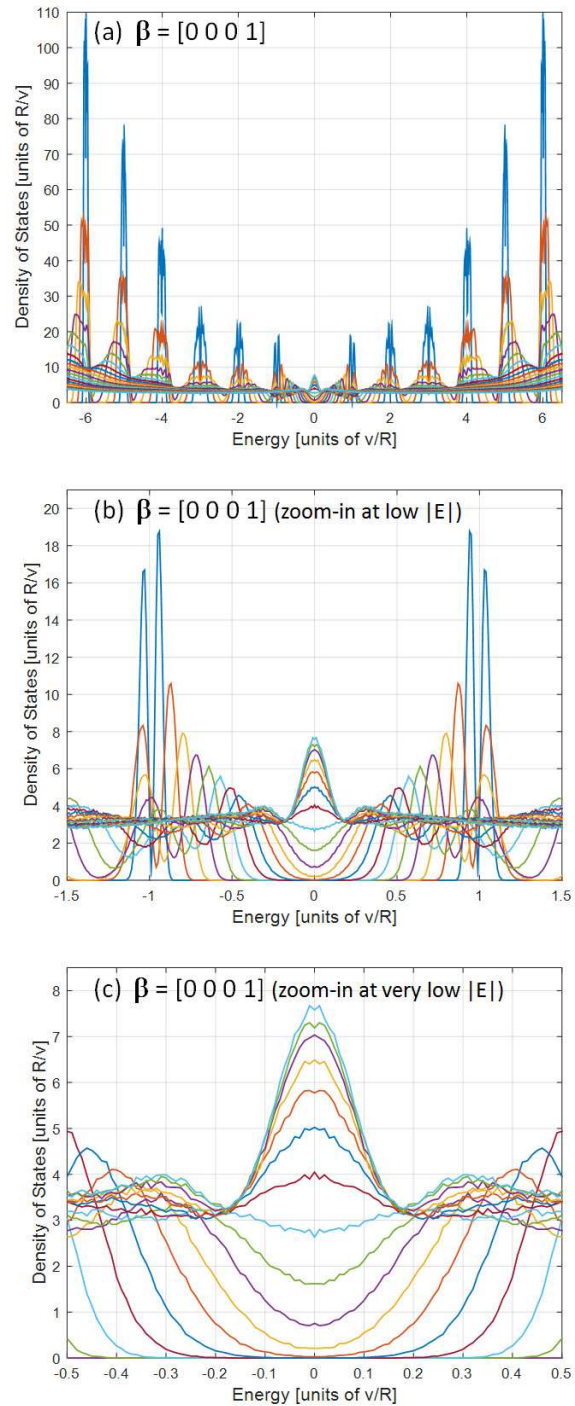}}}
\caption{(Color online) Disorder-averaged density of states for a spherical TI surface in the presence of disorder of type $\beta = [0\ 0\ 0\ 1]$, out-of-surface time-reversal-breaking (spin-dependent) disorder, of nonzero variance only in the $V^3$ direction.  Results for twenty values of disorder strength are plotted in (a), from $V_{\rm scale}=0.2 v/R$ through $4 v/R$ in steps of $0.2 v/R$ (bottom to top at $E=0$).  A zoomed-in view of the same data, for $|E| < 1.5 v/R$, is shown in (b), and a further zoomed-in view, for $|E| < 0.5 v/R$, is shown in (c).}
\label{fig:dadosZ}
\end{figure}

A very different effect is seen for purely out-of-surface time-reversal-breaking disorder (${\bf V}$ parallel to the surface normal).  Here the disorder potential has the form $V = V^3 \sigma_3$, with all other components set to zero, the case we refer to as $\beta = [0\ 0\ 0\ 1]$.  Results are shown in Fig.~\ref{fig:dadosZ}.  Far from the low energy regime, the structure of the disorder-averaged density of states is quite similar to that of the prior case -- split peaks diminishing, broadening, and overlapping as disorder strength increases.  But zooming in on the vicinity of the Dirac point (see Fig.~\ref{fig:dadosZ}(c)), we observe, not a valley, but a clear peak in $\rho(E)$ about $E=0$ that develops as disorder strength grows.  The origin of this feature, and why the out-of-surface and in-surface cases differ so dramatically at low $|E|$, is discussed in Sec.~\ref{sec:explanation}.

\begin{figure}
\centerline{\resizebox{3in}{!}{\includegraphics{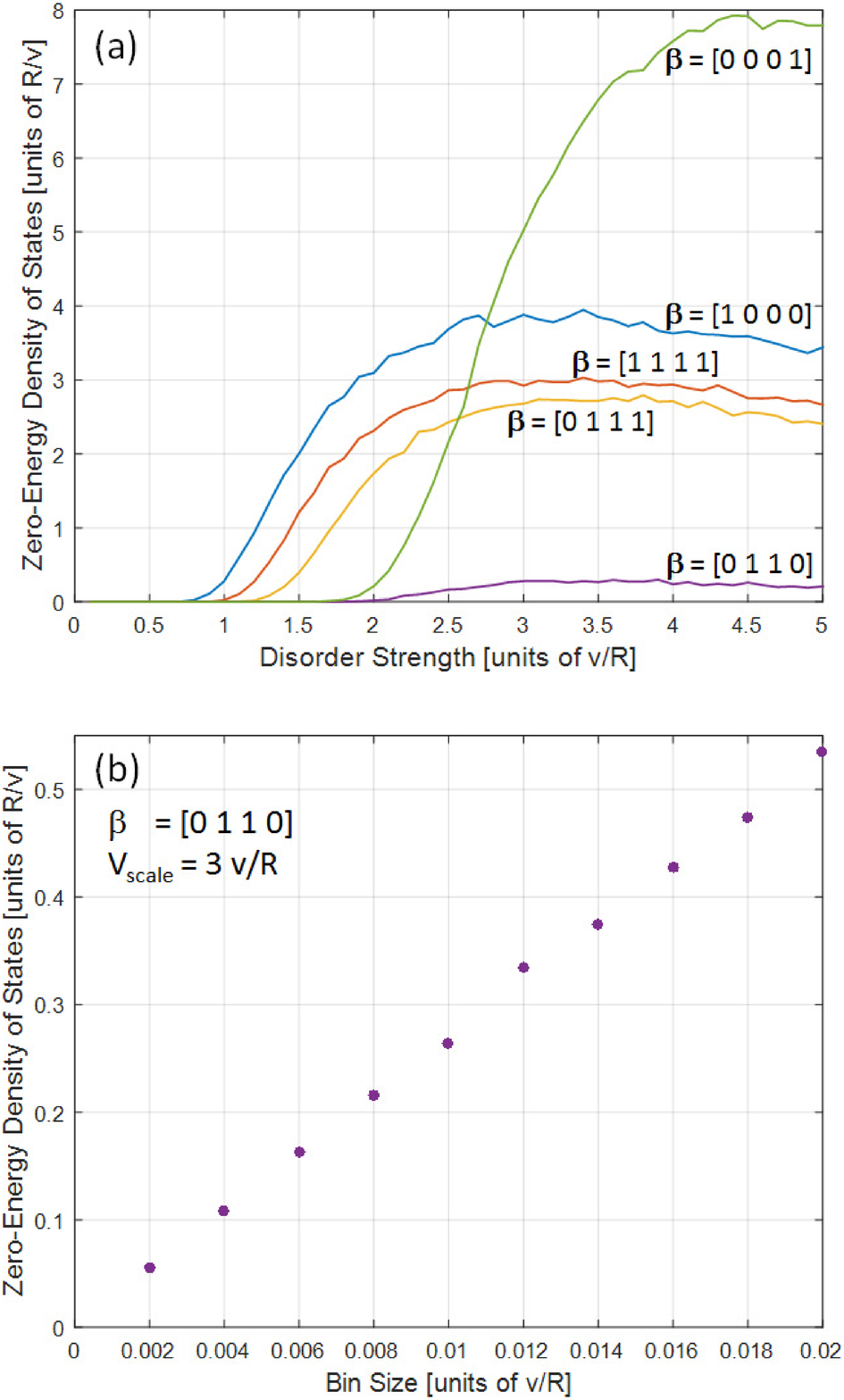}}}
\caption{(Color online) Zero-energy disorder-averaged density of states, $\rho(0)$, for each of the disorder types shown in Figs.~\ref{fig:dadosS} through \ref{fig:dadosZ}.  Plotted in (a) is $\rho(0)$ as a function of disorder strength, $V_{\rm scale}$.  The small but nonzero values attained in the $\beta = [0\ 1\ 1\ 0]$ case are artifacts of averaging over a zero-energy bin of nonzero size, $\Delta E = 0.01 v/R$.  That $\rho(0)$ is strictly zero for this case is clear from panel (b), where the $V_{\rm scale} = 3 v/R$ value is plotted as a function of bin size and seen to linearly approach zero as bin size is decreased (while proportionally increasing the number of disorder instantiations).}
\label{fig:zeroEdados}
\end{figure}

It is instructive to look more closely at the zero-energy value of the disorder-averaged density of states, $\rho(0)$, for each of the five disorder types described above.  These are plotted as a function of disorder strength, $V_{\rm scale}$, in Fig.~\ref{fig:zeroEdados}(a).  In the time-reversal-invariant case ($\beta = [1\ 0\ 0\ 0]$), the time-reversal-breaking with unpolarized ${\bf V}$ case ($\beta = [0\ 1\ 1\ 1]$), and the fully mixed case ($\beta = [1\ 1\ 1\ 1]$), $\rho(0)$ becomes nonzero for sufficient disorder strength ($V_{\rm scale} \sim v/R$) and grows with increasing $V_{\rm scale}$ until saturating at a near-constant value.  In the time-reversal-breaking with out-of-surface ${\bf V}$ case ($\beta = [0\ 0\ 0\ 1]$), $\rho(0)$ grows for approximately twice as long and to approximately twice as large a saturation value (the $E=0$ peak).  Yet in the time-reversal-breaking with in-surface ${\bf V}$ case ($\beta = [0\ 1\ 1\ 0]$), $\rho(0)$ remains zero over the full range of disorder strength.  [Note that a small nonzero value is actually seen in the plot.  This is an artifact of averaging the linear $\rho(E) \sim |E|$ over a nonzero bin size ($\Delta E = 0.01 v/R$) about zero energy to obtain $\rho(0)$.  As shown in Fig.~\ref{fig:zeroEdados}(b), this small value decreases linearly to zero as the bin size is decreased toward zero while the number of disorder instantiations is proportionally increased.]

\begin{figure}
\centerline{\resizebox{3in}{!}{\includegraphics{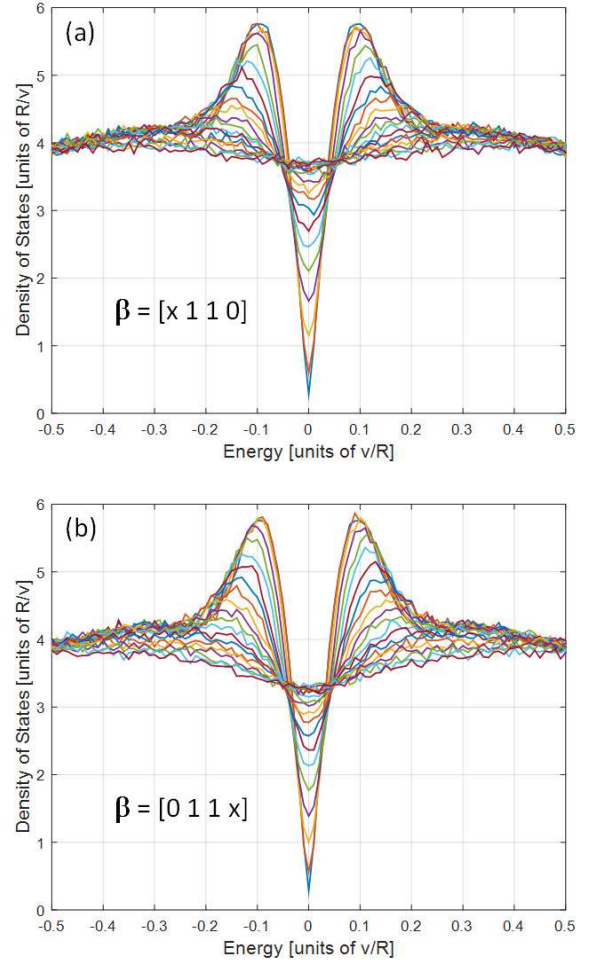}}}
\caption{(Color online) Disorder-averaged density of states for fixed disorder strength, $V_{\rm scale} = 3 v/R$, and slight deviations from disorder type $\beta = [0\ 1\ 1\ 0]$.  Plotted in (a) is disorder of type $\beta = [x\ 1\ 1\ 0]$ for twenty values of $x$ from 0 through 0.2 in steps of 0.01 (bottom to top at $E=0$).  Equivalent plots for $\beta = [0\ 1\ 1\ x]$ are shown in (b).}
\label{fig:dadosXYplusSorZ}
\end{figure}

Finally, let us consider the effect of mixing in a small amount of disorder of a different type on the low-energy linear feature observed for the $\beta = [0\ 1\ 1\ 0]$ case.  In Fig.~\ref{fig:dadosXYplusSorZ}(a), we plot low-energy $\rho(E)$ for $\beta = [x\ 1\ 1\ 0]$ where $x$ varies from 0 through 0.2 in steps of 0.01.  We see that the effect of adding this small $V^0 \openone$ term is to gradually fill in the linear ``hole'' about $E=0$ while diminishing the peaks on either side.  By the time $x=0.2$, the low-energy linear feature is gone.  Mixing in a small $V^3 \sigma_3$ term has a very similar effect, as is seen in Fig.~\ref{fig:dadosXYplusSorZ}(b), where we plot low-energy $\rho(E)$ for $\beta = [0\ 1\ 1\ x]$.  Once again, $x=0.2$ is sufficient to remove the feature entirely.

\section{Explanation of Results}
\label{sec:explanation}
The most intriguing results of this analysis are: (1) the peak splitting observed for purely time-reversal-breaking disorder of the form $V = {\bf V} \cdot {\bf \sigma}$, (2) the low-energy linear feature observed when ${\bf V}$ is oriented in-surface, and (3) the zero-energy peak observed when ${\bf V}$ is oriented out-of-surface. All three of these results can be understood in terms of the symmetries of the Hamiltonian, $H = H_0 + V$, in the presence of different types of disorder.

To begin, note that the clean Hamiltonian, $H_0$ [see Eqs.~(\ref{eq:sphericalDirac}) and (\ref{eq:monopole})], commutes with the time reversal operator ($T \equiv -i\sigma_2 K$ where $K$ denotes complex conjugation) and anticommutes with both $\sigma_3$ and $\sigma_3 T$.
\begin{equation}
\left[ H_0, T \right] = 0 \;\;\;\;\;
\left\{ H_0, \sigma_3 \right\} = 0 \;\;\;\;\;
\left\{ H_0, \sigma_3 T \right\} = 0
\label{eq:H0symmetries}
\end{equation}
The first of these relations reflects the time reversal invariance of $H_0$ and requires the degeneracy of each state with its time-reversed partner.  This must be so because if $\psi$ is an eigenstate of $H$ with eigenvalue $E$ and $[H,T]=0$ then
\begin{equation}
H(T\psi) = T H \psi = T E \psi = E(T\psi)
\label{eq:Tpairdegenderivation}
\end{equation}
so $T\psi$ must also be an eigenstate of $H$ with the same eigenvalue $E$.  The second and third relations in Eq.~(\ref{eq:H0symmetries}) denote two chiral symmetries \cite{bha14} of $H_0$, both of which require that states come in $\pm E$ pairs.  This is the case as long as there exists any operator $Q$ that anticommutes with the Hamiltonian.  To see this, note that if $\psi_+$ is an eigenstate of $H$ with eigenvalue $E$ and $\{H,Q\}=0$ then
\begin{equation}
H(Q\psi_+) = -QH\psi_+ = -QE\psi_+ = -E(Q\psi_+)
\label{eq:pmEderivation}
\end{equation}
so $\psi_- = Q\psi_+$ is an eigenstate of $H$ with eigenvalue $-E$.  Hence, the existence of a chiral symmetry requires that all eigenstates come in $\pm E$ pairs.

Now let us us write down relations corresponding to those of Eq.~(\ref{eq:H0symmetries}) for each of the four terms that can appear in the disorder potential.
\begin{equation}
\begin{array}{rrr}
\left[ V^0 \openone, T \right] = 0 \vspace{0.2cm} & \left[ V^0 \openone, \sigma_3 \right] = 0 & \left[ V^0 \openone, \sigma_3 T \right] = 0 \\
\left\{ V^1 \sigma_1, T \right\} = 0 \vspace{0.2cm} & \left\{ V^1 \sigma_1, \sigma_3 \right\} = 0 & \left[ V^1 \sigma_1, \sigma_3 T \right] = 0 \\
\left\{ V^2 \sigma_2, T \right\} = 0 \vspace{0.2cm} & \left\{ V^2 \sigma_2, \sigma_3 \right\} = 0 & \left[ V^2 \sigma_2, \sigma_3 T \right] = 0 \\
\left\{ V^3 \sigma_3, T \right\} = 0 \vspace{0.2cm} & \left[ V^3 \sigma_3, \sigma_3 \right] = 0 & \left\{ V^3 \sigma_3, \sigma_3 T \right\} = 0
\end{array}
\label{eq:Valphasymmetries}
\end{equation}
These follow from the realness of the $V^\alpha$ functions and the fact that each Pauli matrix commutes with itself but anticommutes with the other two Pauli matrices.  Thus, while $H_0$ commutes with $T$ and anticommutes with $\sigma_3$ and $\sigma_3 T$, the addition of any one of the four possible disorder potential terms breaks two of these three symmetries.

Consider the $\beta = [1\ 0\ 0\ 0]$ case.  For each instantiation of the disorder, the full Hamiltonian, $H_{1000} = H_0 + V^0 \openone$, preserves time reversal symmetry but breaks both chiral symmetries.
\begin{equation}
\left[ H_{1000}, T \right] = 0 \;\;\;\;\;
\left\{ H_{1000}, \sigma_3 \right\} \neq 0 \;\;\;\;\;
\left\{ H_{1000}, \sigma_3 T \right\} \neq 0
\label{eq:H1000symmetries}
\end{equation}
Thus, all eigenstates are doubly degenerate (time-reversed pairs) but they no longer come in $\pm E$ pairs, so there is nothing special about $E=0$ and $\rho(E)$ exhibits no special feature at low energy.

In the $\beta = [0\ 1\ 1\ 1]$ case, the full Hamiltonian, $H_{0111} = H_0 + V^1 \sigma_1 + V^2 \sigma_2 + V^3 \sigma_3$, breaks all three symmetries.
\begin{equation}
\left[ H_{0111}, T \right] \neq 0 \;\;\;\;\;
\left\{ H_{0111}, \sigma_3 \right\} \neq 0 \;\;\;\;\;
\left\{ H_{0111}, \sigma_3 T \right\} \neq 0
\label{eq:H0111symmetries}
\end{equation}
With both chiral symmetries broken, the eigenstates, once again, do not come in $\pm E$ pairs, so there is again nothing special about $E=0$ and no special feature in $\rho(E)$ at low energy.  Furthermore, since the disorder potential, $V$, {\it anticommutes} with $T$ while $H_0$ commutes with $T$, the previously degenerate time-reversed-pair states of the clean Hamiltonian are shifted oppositely by weak disorder.  This is so because
\begin{equation}
\langle T\psi | V | T\psi \rangle = - \langle T\psi | T (V\psi) \rangle = - \langle V\psi | \psi \rangle = - \langle \psi | V | \psi \rangle
\label{eq:oppositeshifts}
\end{equation}
where the first equality follows because $\{V,T\}=0$, the second because $T$ is antiunitary, and the third because $V$ is hermitian.  This is why we see peak splitting in the present case, as well as all other cases where $\{V,T\}=0$, including $\beta = [0\ 1\ 1\ 0]$ and $\beta = [0\ 0\ 0\ 1]$, but not for $\beta = [1\ 0\ 0\ 0]$ which preserves time-reversal invariance.  For $\beta = [1\ 1\ 1\ 1]$, the ${\bf V} \cdot {\bf \sigma}$ term anticommutes with $T$ and shifts time-reversed-pair states oppositely while the $V^0 \openone$ term commutes with $T$ and provides the same energy shift to both members of each pair.  The peak splitting effect is therefore muted.

Now let us consider the origin of the low-energy (Dirac-point-vicinity) features seen for in-surface and out-of-surface time-reversal-breaking disorder.  An important clue can be found in the structure of the {\it noise} in the $\rho(E)$ vs.\ $E$ plots for $\beta = [0\ 1\ 1\ 0]$ and $\beta = [0\ 0\ 0\ 1]$.  Despite averaging over 200,000 disorder instantiations, there is always some noise in our data because we have not included an infinite number of instantiations in our average.  Careful inspection of each set of plots (Figs.~\ref{fig:dadosS} through \ref{fig:dadosZ}) reveals that the noise is an even function of energy for $\beta = [0\ 1\ 1\ 0]$ and $\beta = [0\ 0\ 0\ 1]$ but exhibits no identifiable symmetry in the other three cases.

To see why, note that for $\beta = [0\ 1\ 1\ 0]$, the full Hamiltonian, $H_{0110} = H_0 + V^1 \sigma_1 + V^2 \sigma_2$, breaks the second chiral symmetry, but not the first.
\begin{equation}
\left[ H_{0110}, T \right] \neq 0 \;\;\;\;\;
\left\{ H_{0110}, \sigma_3 \right\} = 0 \;\;\;\;\;
\left\{ H_{0110}, \sigma_3 T \right\} \neq 0
\label{eq:H0110symmetries}
\end{equation}
And for $\beta = [0\ 0\ 0\ 1]$, $H_{0001} = H_0 + V^3 \sigma_3$ breaks the first chiral symmetry, but not the second.
\begin{equation}
\left[ H_{0001}, T \right] \neq 0 \;\;\;\;\;
\left\{ H_{0001}, \sigma_3 \right\} \neq 0 \;\;\;\;\;
\left\{ H_{0001}, \sigma_3 T \right\} = 0
\label{eq:H0001symmetries}
\end{equation}
Therefore, in both cases, there exists an operator ($\sigma_3$ for the former case and $\sigma_3 T$ for the latter) that anticommutes with the full Hamiltonian, so all eigenstates come in $\pm E$ pairs.  As a result, the energy spectrum for every instantiation of disorder is even in energy, so even the noise in $\rho(E)$ must be an even function of $E$, as is clear from Figs.~\ref{fig:dadosXY} and \ref{fig:dadosZ}.  In all the other cases, where both chiral symmetries are broken, states do not come in $\pm E$ pairs so the noise need not satisfy any particular symmetry.

When eigenstates come in $\pm E$ pairs, $E=0$ is quite special.  It is the energy at which eigenstates, when driven to low energy by disorder, can meet their partners and, possibly, hybridize with them.  But note that the partner states are different for the in-surface versus out-of-surface cases because it is a different chiral symmetry that survives for each case.  If $\psi_+$ is a positive energy eigenstate, its negative energy partner is $\sigma_3 \psi_+$ for the in-surface case and $\sigma_3 T \psi_+$ for the out-of-surface case.  The fact that different low-energy density-of-states features are observed in each case is a direct result of this difference in $\pm E$ partners.

Consider what happens, in either case, when disorder is strong enough to bring $\pm E$ pairs close to zero energy.  Let $\Delta V$ be the small additional perturbation that would, in the absense of hybridization, make $\psi_+$ and $\psi_-$ degenerate at $E=0$.  It is then a simple matter of two-state degenerate perturbation theory to see if the partners hybridize or not.  We need only calculate the off-diagonal matrix element
\begin{equation}
\Delta V_{+-} = \langle \psi_+ | \Delta V | \psi_- \rangle .
\label{eq:DeltaVpmdef}
\end{equation}
Note that $\psi_+$ and $\psi_-$ no longer resemble the clean eigenstates, which have already been thoroughly mixed by the disorder potential.  Thus, we do not know the unperturbed states, only their relation to each other.  But it turns out that is enough.

For the in-surface $\beta = [0\ 1\ 1\ 0]$ case,
\begin{equation}
\Delta V = \Delta V_1(\theta,\phi) \sigma_1 + \Delta V_2(\theta,\phi) \sigma_2
\label{eq:DeltaV0110}
\end{equation}
and since $\{H_{0110},\sigma_3\}=0$, we can write
\begin{equation}
\psi_+ \equiv \left[ \begin{array}{c} u \\ v \end{array} \right] \;\;\;\;\;\;
\psi_- = \sigma_3 \psi_+ = \left[ \begin{array}{c} u \\ -v \end{array} \right]
\label{eq:psipm0110}
\end{equation}
where $u$ and $v$ are complex scalar functions of $\theta$ and $\phi$.  Plugging into Eq.~\ref{eq:DeltaVpmdef} then reveals that
\begin{equation}
\Delta V_{+-} = 2i \int_0^{2\pi} \!\!\!\!\!\!d\phi \int_0^\pi \!\!\!\!d\theta \sin\theta
\left[ \Delta V_2 \mbox{Re}(u^* v) - \Delta V_1 \mbox{Im}(u^* v) \right]
\label{eq:DeltaVpm0110}
\end{equation}
Since this is generally nonzero, the $\pm E$ partner states do hybridize, resulting in a level repulsion that yields a lack of states about E=0.  Hence, as we increase disorder strength, partner states anti-cross at zero energy, with anti-crossing gaps of varying size.  This happens generically, for every disorder instantiation that drives states toward low energy, which explains why we find $\rho(0) = 0$ for disorder of this type.  Prevented, by the chiral symmetry, from reaching zero energy, states driven toward low energy build up in two symmetric peaks, with density of states falling to zero as energy goes to zero.  Hence the low-energy linear feature seen in Fig.~\ref{fig:dadosXY}(c).

But why doesn't the same thing happen for the out-of-surface $\beta = [0\ 0\ 0\ 1]$ case?  The scenario of $\pm E$ partner states driven toward each other by the disorder potential is precisely the same as above, only now
\begin{equation}
\Delta V = \Delta V_3(\theta,\phi) \sigma_3
\label{eq:DeltaV0001}
\end{equation}
and it is the other chiral symmetry that is preserved: $\{H_{0001},\sigma_3 T\}=0$.  So we write
\begin{equation}
\psi_+ \equiv \left[ \begin{array}{c} u \\ v \end{array} \right] \;\;\;\;\;\;
\psi_- = \sigma_3 T \psi_+ = -\sigma_1 K \psi_+ = \left[ \begin{array}{c} -v^* \\ -u^* \end{array} \right]
\label{eq:psipm0001}
\end{equation}
and plug into Eq.~\ref{eq:DeltaVpmdef}, just as before.  But in this case, we find a very different result:
\begin{equation}
\Delta V_{+-} = \int_0^{2\pi} \!\!\!\!\!\!d\phi \int_0^\pi \!\!\!\!d\theta \sin\theta
\left[ \Delta V_3 \left( -u^* v^* + u^* v^* \right) \right] = 0 .
\label{eq:DeltaVpm0001}
\end{equation}
The off-diagonal matrix element is zero for {\it any} functions $u$, $v$, and $\Delta V_3$.  The $\pm E$ partner states therefore do not hybridize when they meet at zero energy, there is no level repulsion, and states about $E=0$ are plentiful.  As we increase disorder strength, partner states simply cross at zero energy.  Zero energy is once again special, but now for the opposite reason.  This coming together of states from positive and negative energy yields an enhancement in the density of states, resulting in a peak about $E=0$.  Hence the zero-energy peak seen in Fig.~\ref{fig:dadosZ}(c).

\section{Conclusions}
\label{sec:conclusions}
In this paper, we have considered a topological insulator (TI) of spherical geometry and studied the influence of disorder on the density of surface states.  This spherical TI problem is relevant both to the study of physical systems of spherical geometry (i.e.\ TI nanoparticles) and as a theoretical construct with the purpose of providing insight regarding the physics of the flat TI surface that is recovered in the large radius limit.  A great advantage of this construct is that the finite size of the spherical surface results in a discrete energy spectrum with well-defined eigenstates that is well suited to numerical analysis.  The eigenvalues and eigenstates of the clean TI surface Hamiltonian, $H_0$, elegantly derived and developed in Refs.~\onlinecite{neu15,imu12,gre11,hal83}, provided the basis for our analysis.  To $H_0$ we added a disorder potential of the most general hermitian form, $V(\theta,\phi) = V^0(\theta,\phi) \openone + {\bf V}(\theta,\phi) \cdot {\bf \sigma}$, where $V_0$ represents time-reversal-invariant (spin-independent) disorder and the three-vector ${\bf V}$ represents time-reversal-breaking (spin-dependent) disorder.  Expanding the four disorder functions in spherical harmonics, we randomly drew coefficients from a gaussian distribution in four-dimensional parameter space.  We then evaluated the full Hamiltonian, $H = H_0 + V$, in the basis of the clean eigenstates, diagonalized to find the energy spectrum, binned the results to obtain the density of states, and averaged over 200,000 disorder instantiations to compute the disorder-averaged density of states, $\rho(E)$.  We considered disorder of varying strength ($V_{\rm scale}$) and of different types (see $\beta$-vector nomenclature introduced in Sec.~\ref{ssec:random}) by controlling the $4 \times 4$ covariance matrix of our zero-mean gaussian distribution.

Quite generically, we find that increasing disorder strength leads to the broadening, decay, and overlap of the Landau level peaks that characterize the clean density of states.  We observed this scenario for all the different disorder types (characterized by different $\beta$-vectors) that we considered.  Such effects are to be expected as the disorder potential thoroughly mixes the eigenstates of the clean Hamiltonian.  However, our results also revealed striking differences, rooted in the symmetries of the TI surface Hamiltonian, between density-of-states functions calculated for different types of disorder.

We found that the broadened peak structure characteristic of the weak-disorder regime manifests quite differently for spin-dependent disorder (only the ${\bf V} \cdot {\bf \sigma}$ terms) than it does for spin-independent disorder (only the $V^0 \openone$ term).  The former case results in a distinct splitting of the Landau level peaks (see Fig.~\ref{fig:dadosXYZ}) that is absent in the latter (see Fig.~\ref{fig:dadosS}).  As discussed in Sec.~\ref{sec:explanation}, this is a consequence of the time-reversal invariance of the clean Hamiltonian and the fact that the spin-dependent disorder potential terms anticommute with the time-reversal operator.

\begin{figure}
\centerline{\resizebox{3in}{!}{\includegraphics{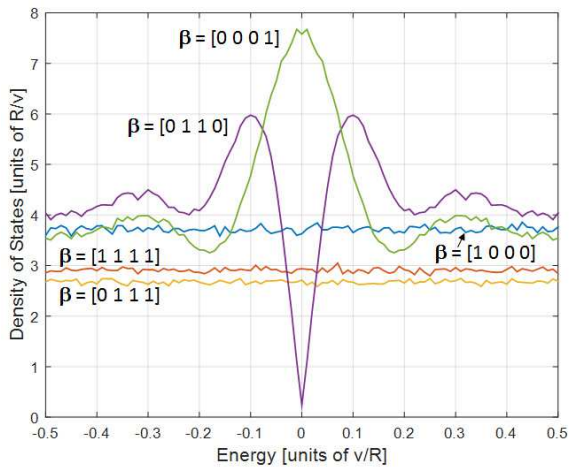}}}
\caption{(Color online) Low-energy disorder-averaged density of states for strong disorder of fixed strength, $V_{\rm scale} = 4 v/R$, for each of the disorder types shown in Figs.~\ref{fig:dadosS} through \ref{fig:dadosZ}.}
\label{fig:dadosFixedVscale}
\end{figure}

Furthermore, our calculations revealed significant, disorder-type-dependent differences in the structure of the strong-disorder density of states in the vicinity of zero energy.  For most types of disorder, increasing disorder strength pushes states closer and closer to zero energy, resulting in a low-energy density of states that becomes nonzero for sufficient disorder and continues to increase until saturating at an energy-independent value.  This is what we see for $\beta = [1\ 0\ 0\ 0]$, $\beta = [1\ 1\ 1\ 1]$, and $\beta = [0\ 1\ 1\ 1]$.  But something very different happens for spin-dependent disorder with ${\bf V}$ either entirely in-surface ($\beta = [0\ 1\ 1\ 0]$) or entirely out-of-surface ($\beta = [0\ 0\ 0\ 1]$), as can be seen in Fig.~\ref{fig:dadosFixedVscale}.  In the in-surface case, increasing disorder strength yields an enhancement of the density of states {\it near} zero energy but not {\it at} zero energy.  The zero-energy density of states is always strictly zero, and we see a linear increase with increasing $|E|$, up to peaks in the vicinity of $E \approx \pm 0.1 v/R$.  By contrast, in the out-of-surface case, increasing disorder strength yields a peak at zero energy, reaching a maximum value of roughly twice the constant density of states seen at higher energies.  As discussed in Sec.~\ref{sec:explanation}, what makes these two cases so special is that for each, the full Hamiltonian, $H = H_0 + V$, preserves one of the two chiral symmetries of the clean Hamiltonian, $H_0$.  As a result, eigenvalues of $H$ come in $\pm E$ pairs, so the density of states for each individual disorder instantiation is even in $E$.  For this reason, the disorder-averaged density of states, $\rho(E)$, for these two cases (and not the others) are strictly even functions of energy, down to the noise (see Fig.~\ref{fig:dadosFixedVscale}).  Also for this reason, $E=0$ is quite special in both of these cases, for it is the energy where $\pm E$ partner states meet each other, given sufficient disorder.  But what makes these two cases so different from each other, is that for each, it is a {\it different} chiral symmetry of the clean Hamiltonian that is preserved by the full Hamiltonian.  In the in-surface case, $\{H,\sigma_3\}=0$, so the $\pm E$ partner states are $\psi$ and $\sigma_3 \psi$, and we find that these two states hybridize when they meet, resulting in an anti-crossing (level repulsion), and therefore a depletion of states at zero energy.  In the out-of-surface case, $\{H,\sigma_3 T\}=0$, so the $\pm E$ partner states are $\psi$ and $\sigma_3 T \psi$, and we find that these two states {\it do not} hybridize when they meet, so there is no level repulsion, and therefore an abundance of states at zero energy.

We expect that the disorder effects discussed herein should be relevant to experiment, including the spectroscopy of topological insulator nanoparticles, where an ensemble average over many nanoparticles would play the role of our disorder average.  Most observable would be the broadening, decay, and overlap of Landau level peaks that we see generically for all types of disorder.  More challenging would be the observation of the weak-disorder peak splitting that is characteristic of time-reversal-breaking disorder.  And most challenging would be the observation of the low-energy (Dirac-point-vicinity) density-of-states features that we predict for strong, time-reversal-breaking disorder of the in-surface ($\beta = [0\ 1\ 1\ 0]$) and out-of-surface ($\beta = [0\ 0\ 0\ 1]$) types, as the requisite control over the form of the disorder potential may be difficult to achieve.

In future theoretical work, we intend to pursue a synthesis of this disorder analysis with the interactions study performed on the same system by Neupert {\it et al.\/} \cite{neu15} in order to probe the interplay between interactions and disorder.

\begin{acknowledgments}
I am grateful to S. Ganeshan for introducing me to the work of Neupert {\it et al.\/} \cite{neu15} and to B. Burrington and G. C. Levine for very helpful discussions.  This work was supported by funds provided by Hofstra University, including a Faculty Research and Development Grant (FRDG), a Presidential Research Award Program (PRAP) grant, and faculty startup funds.
\end{acknowledgments}

\end{document}